\documentstyle[12pt]{article}
\let\section=\subsection     \let\subsection=\subsubsection                
                           
\def\sun{$SU(N_c)\;$}
\begin{document}
\begin{center}
{\large \bf WHY THE $SU(\infty)$ DECONFINING PHASE}\\[2mm]
{\large \bf TRANSITION MIGHT BE OF SECOND ORDER}
\footnote{{\it To appear in the proceedings of the XXV Hirschegg Workshop 
on "QCD Phase Transitions", Jan. 1997.}}
\\[5mm]
ROBERT D.~PISARSKI AND MICHEL~TYTGAT \\[5mm]
{\small \it  Dept. of Physics, Brookhaven National Lab\\ 
Upton, NY 11973 USA\\[8mm] }
\end{center}
\begin{abstract}\noindent
Based upon what is known about the phase transition(s)
of an $SU(3)$ gauge theory,
we argue that in a $SU(N_c)$ gauge theory without quarks,
at nonzero temperature the deconfining phase transition 
is of second order when $N_c \geq 4$.
\end{abstract}

From 't Hooft, Witten, and others,
it is known that at zero temperature,
many properties of $QCD$ can be understood
by assuming that for a $SU(N_c)$ gauge theory, 
$N_c=3$ is reasonably well described 
by the limit of $N_c = \infty$  \cite{large}.
This includes such features as the OZI rule, the
approximate mass degeneracy between the $\rho$ and $\omega$ vector
mesons, {\it etc.} \cite{teper}.

In this note we suggest that the large $N_c$ expansion
can also provide a way of understanding the phase diagram
of a $SU(3)$ gauge theory 
at nonzero temperature {\it if} the phase transition
in a ``pure'' \sun gauge theory (without dynamical
quark fields) is of {\it second} order whenever 
$N_c \geq 4$, including $N_c = \infty$.

Simply counting the number of degrees of freedom allows
one to make extremely strong statements about the
thermodynamics of an \sun gauge 
theory \cite{thorn}-\cite{mclerran}.  The basic
point is simply that because gluons are in the adjoint
representation, and quarks in the fundamental, at
large $N_c$ what happens to the $\sim N_c^2$ gluons
totally dominates the $\sim N_c$ quarks.  In the confined
phase, confinement implies that all states are bound
into colorless hadrons, so that the free energy is
of necessity $\sim 1$.  At high temperature, the
free energy is expected to be $\sim N_c^2$.  Thus,
as pointed out first by Thorn \cite{thorn,pisarski}, one can
use the free energy itself to define the deconfining
phase transition,
\begin{equation}
F(T) \sim 1 \;\;\; , \;\;\; T \leq T_d \;\;\; ; \;\;\;
F(T) \sim N_c^2 \;\;\; , \;\;\; T \geq T_d \; .
\label{eq:1}
\end{equation}
(See, however, \cite{mclerran}.)
In general, the deconfining phase transition is
rigorously related to the global $Z(N_c)$ symmetry
of a \sun gauge theory, where the order parameter is
the Wilson line \cite{yaffe,reviews},
\begin{equation}
L(\vec{x}) = tr \left({\cal P} {\rm exp} 
\left( i g \int^{1/T}_0 A_0(\vec{x},\tau) d\tau \right)\right) \; .
\label{eq:2}
\end{equation}
We assume that the $Z(N_c)$ symmetry is
broken above $T_d$, $\langle L \rangle = 0 $ for $T \leq T_d$,
$\langle L \rangle \neq 0 $ for $T \geq T_d$,
which is most reasonable.

For $n_f$ flavors of massless quarks,
the analysis of the chiral phase transition proceeds
as usual \cite{chiral}.  The only difference is
that since the effects of the axial anomaly
are $\sim g^2 n_f \sim (g^2 N_c) n_f/N_c$, if $n_f$ and
$g^2 N_c$ are held fixed as $N_c \rightarrow \infty$, then
the effects of the anomaly vanish, and the global chiral symmetry is 
$S(U(n_f) \times U(n_f))$.  We assume this is
broken to $SU(n_f)$ at zero temperature \cite{large},
and restored at a temperature $T_\chi$.  Whatever
the order of the chiral transition, however, since
that part of the free energy is again $\sim N_c$, it
cannot affect $T_d$.  Thus at infinite $N_c$ we can
precisely characterize both the chiral and deconfining
phase transitions.  For the purposes of argument we take
$T_\chi = T_d$.

Suppose now that the deconfining phase transition is of
{\it first} order.  Whatever the nature of the chiral
transition, if we hold $n_f$ finite as $N_c \rightarrow \infty$,
the gluons dominate, and for any number of quark flavors,
the first order deconfining transition always wins.
Near $T_d$, the effective three dimensional theory for
$L(\vec{x})$ is
\begin{equation}
{\cal L} = \frac{1}{2} |\partial_\alpha L|^2
+ \frac{1}{2} m^2 |L|^2 + g_4 (|L|^2)^2
+ g_6 (|L|^2)^3 \; . 
\label{eq:3}
\end{equation}
A first order transition implies that the quartic coupling
is negative, $g_4 < 0$.  This is possible because the most
general renormalizable theory in three dimensions includes
a six-point coupling, which for stability must be positive,
$g_6 > 0$.

At present, numerical simulations of an $SU(3)$ gauge theory demonstrate
the following \cite{lattice}.
In the pure gauge theory, if $\epsilon$ is the energy density,
the latent heat is relatively weak \cite{karsch}, 
\begin{equation}
\left. \frac{\Delta \epsilon}{\epsilon}\right|_{T_d} \leq
\frac{1}{3} \; .
\label{eq:4}
\end{equation}
If dynamical fermions are added, the deconfining
transition can be washed out entirely; in the ``Columbia'' phase
diagram, as a function of $m_u = m_d$ and $m_s$, there is
a clean gap seperating the regions of a first order deconfining
phase transition, for large quark masses, and a first
order chiral phase transition, for small quark masses \cite{lattice}.

These features of the $N_c=3$ phase diagram are difficult
to understand if the large $N_c$ expansion 
is a reasonable guide, and if the 
deconfining phase transition for $N_c = \infty$
is strongly first order.  One would expect that the deconfining
phase transition would be strongly first order at $N_c = 3$,
and that the deconfining transition would dominate for all quark
masses.  Thus there would be no need to draw the Columbia phase
diagram --- the transition would always be first order.

On the other hand, if the deconfining phase transition is
of {\it second} order, then at least in a handwaving sense, everything
seems to fit.  Without dynamical fermions, at $N_c = 3$
the latent heat is small, down by $\sim 1/N_c \sim 1/3$.
Further, since the deconfining phase transition is relatively
weak to begin with, it is easily washed out by dynamical quarks.
The chiral transition is of first order for three massless flavors,
and about that point, but that is special to the chiral transition.

Moreover, the large $N_c$ expansion does provide an understanding
of one very familiar feature of the phase transition in an $SU(3)$ gauge
theory: the large increase in entropy.  This is due, naturally,
to the vast increase in the number of degrees of freedom between
the hadronic and deconfined phases.  But this terminology only
makes sense if we can speak of a deconfined phase.  Why there
is such a large increase in entropy for small quark masses,
when the transition is manifestly dominated by the chiral properties?
No effective model of the chiral transition will produce such
a large jump in entropy, simply because there is no great
change in the number of (light) degrees of freedom.  If we
think of a second order transition for $N_c = \infty$, though,
we automatically get a large increase in entropy.  Not a jump,
just an smooth but sharp increase.

We acknowledge that our suggestion contradicts 
known results from lattice gauge theory.  Using the Eguchi-Kawai
approximation to large $N_c$ \cite{eguchi}, under the assumption
that the coupling between spacelike plaquettes can be neglected,
Gocksch and Neri proved that the deconfining phase transition is
of first order \cite{gocksch,more}.  
(See, however, \cite{klink}.)  We note that a different
approach to large $N_c$ by Yaffe {\it et al.} \cite{yaffe1}
appears to indicate that the deconfining transition 
is of second order at $N_c = \infty$ \cite{yaffe2}.  

Numerical simulations of a $SU(4)$ lattice
gauge theory have been done \cite{mc4}, 
and indicate a first order deconfining
phase transition.  Here we can only suggest that perhaps what
was observed is a bulk transition, seperate from the true
deconfining phase transition at nonzero temperature.

In that regards, we
would also like to make a technical aside about
the deconfining phase transition for $N_c = 4$.
For arbitrary
$N_c$, the general effective lagrangian includes (\ref{eq:3}),
which posseses a global $O(2)$ symmetry, and the term
\begin{equation}
{\cal L}_{Z} = g_Z \left(L^{N_c} 
+ \left( L^* \right)^{N_c} \right) \; .
\label{eq:6}
\end{equation}
which is only invariant under global $Z(N_c)$ transformations.
For $N_c = 2$ this changes the symmetry from $O(2)$ to
$Z(2)$.  When $N_c = 3$ this is a cubic coupling, and 
drives the deconfining phase transition
first order \cite{yaffe}.

For $N_c = 4$, the coupling $g_Z$ in ${\cal L}_Z$ is
as important as $g_4$ in ${\cal L}$, and one must be more careful.
In particular, there is the possibility
that having both $g_Z$ and $g_4 \neq 0$ produces a fluctuation
induced first order transition.  For $N_c = 4$,
the theory with the lagrangian ${\cal L} + {\cal L}_Z$ is
equivalent to what is known as the $n=2$ model of 
cubic anisotropy.  In $4-\epsilon$ dimensions,
the $O(2)$ fixed point is infrared stable \cite{cubic}.  
This is supported by recent Monte Carlo simulations directly in 
three dimensions \cite{mc3}.  Hence, for $N_c = 4$,
it does not appear as if the transition is fluctuation induced
first order, and is of second order when $g_4$ and $g_Z$ are positive.
Of course the deconfining phase transition for $SU(4)$
could still be first order because the couplings $g_4$ and/or
$g_Z$ are negative to begin with.  

For $N_c = 5$, one expects that the pentic coupling
can be neglected relative to the quartic.  This must
be qualified: precisely at
the tricritical point, where $m^2 = g_4 = 0$, 
there is only a pentic and a hexic coupling.
If $g_Z \neq 0$, the pentic coupling wins, and because it is 
odd in $\phi$, drives the transition first order.
This remains true in a region around the
tricritical point, such as for $m^2=0$ and small $g_4>0$,
when the dimensionless quantity $g_Z/\sqrt{g_4} \gg 1$. 
If $g_Z/\sqrt{g_4} \ll 1$, however, then the pentic coupling
is negligible relative to the quartic,
and the transition is of second order.  This is in
contrast to $N_c = 3$, where the cubic coupling
$g_Z \neq 0$ drives the transition first order
regardless of the magnitude of $g_Z/|g_4|^{3/2}$:
it is weakly first order for $g_Z/|g_4|^{3/2} \ll 1$,
but strongly first order when $g_Z/|g_4|^{3/2} \gg 1$.
The difference is because the quartic coupling is less
relevant than a cubic, but more relevant than a pentic.

For $N_c \geq 6$, we can certainly neglect the $Z(N_c)$ coupling.
Thus we see that with some technical qualifications, that
the conclusions of \cite{yaffe} remain: for $g_4 \geq 0$,
when $N_c \geq 4$
the $SU(N_c)$ transition is second order, in the universality
class of an $O(2)$ model.

We conclude with two suggestions.

The first is to measure the coupling $g_4$ for $SU(3)$ and
see if it is positive.
For a pure $SU(2)$ gauge theory, it appears as if the
deconfining phase transition is of second order \cite{su2}
(see, however, \cite{gavai}), which 
implies that the coupling $g_4$ is positive.
Thus it would be interesting to know if $g_4 > 0$ for
$SU(3)$; if so, it would be reasonable to assume that
$g_4 > 0$ for all $N_c$; this implies that the deconfining
phase transition is of second order whenever $N_c \geq 4$.

A second suggestion is simply to carefully measure again the deconfining
phase transition for $SU(4)$.  Certainly $N_c = 4$ is closer
to $N_c = \infty$ than $N_c = 3$.  

Of course our arguments are most indirect, with many
obvious loopholes: large $N_c$ may not describe thermodynamics
for $N_c = 3$; the $N_c = \infty$ transition might be weakly
first order (but then - why?); and so on.  Still, a second
order phase transition at $N_c = \infty$ helps one understand
understand many qualitative features of the phase
diagram at $N_c = 3$.

We thank Prof.'s Bill\'o, Caselle, D'Adda, Panzeri, and Yaffe
for discussions on large $N_c$, and
Prof. Rajagopal for a comment on $N_c = 5$.
This work is supported by a DOE grant at 
Brookhaven National Laboratory, DE-AC02-76CH00016.

\end{document}